\documentclass[a4paper,12pt]{article}
\usepackage[utf8]{inputenc}
\usepackage[textwidth=18.0cm,textheight=25cm]{geometry}
\usepackage{authblk}
\usepackage{graphics}
\usepackage{amsfonts,amssymb,amsmath}
\usepackage{graphicx,color,epsf,subfigure}
\usepackage[T1]{fontenc}
\usepackage[utf8]{inputenc}

\newcommand{\x}{\mathbf{x}}

\newcommand{\gb}{\lambda_{_\text{GB}}}

\newcommand{\eq}[1]{(\ref{eq:#1})}  

\title{\bf Expanding plasmas from Anti de Sitter black holes}
\author[]{Giancarlo Camilo\footnote{Email: \texttt{gcamilo@usp.br}}}
\affil[]{\small Instituto de F\'isica, Universidade de S\~ao Paulo\break
Rua do Mat\~ao, Travessa R, 187, Cidade Universit\'aria, 05508-090, S\~ao Paulo, Brasil}

\date{\small (Dated: November 23, 2016)}

\begin{document}

\maketitle

\begin{abstract}
We introduce a new foliation of AdS$_5$ black holes such that the conformal boundary takes the form of a $4$-dimensional FLRW spacetime with scale factor $a(t)$. The foliation employs Eddington-Finkelstein-like coordinates and is applicable to a large class of AdS black holes, supported by matter fields or not, considerably extending previous efforts in the literature. We argue that the holographic dual picture of a CFT plasma on a FLRW background provides an interesting prototype to study the nonequilibrium dynamics of expanding plasmas and use holographic renormalization to extract the renormalized energy-momentum tensor of the dual plasma. We illustrate the procedure for three black holes of interest, namely AdS-Schwarzschild, AdS-Gauss-Bonnet, and AdS-Reissner-Nordström. For the latter, as a by-product, we show that the nonequilibrium dynamics of a CFT plasma subject to a quench in the chemical potential (i.e., a time-dependent chemical potential) resembles a cosmological evolution with the scale factor $a(t)$ being inversely related to the quench profile $\mu(t)$.
\end{abstract}

\section{Introduction}
\label{AdSFLRWintro}

The gauge/gravity duality (see, e.g., \cite{Ammon:2015wua} for a recent review) is currently established as a valuable tool to approach the physics of strongly coupled quantum systems. The general idea is that, by mapping certain quantum systems to an equivalent dual gravity theory living in a higher dimensional bulk spacetime, difficult problems may become tractable. The duality has found a wide range of applications over the last years, specially when it comes to the nonequilibrium dynamics of quantum field theories at strong coupling, which tipically involves time-dependent solutions on the gravity side. Examples include the thermalization of non-Abelian plasmas similar to the quark gluon plasma formed in heavy ion collisions at the RHIC and LHC, which is mapped to an Anti de Sitter (AdS) black hole formation process on the gravity side \cite{Chesler:2008hg,Chesler:2010bi,Balasubramanian:2010ce,Heller:2011ju,Camilo:2014npa,Keegan:2015avk}, and the quench dynamics of quantum systems, where a time-dependent coupling (or \emph{quench}) on the field theory side translates into a boundary condition for bulk fields in the dual gravity description \cite{Das:2011nk,Buchel:2013lla,Buchel:2013gba,Buchel:2014gta,Camilo:2015wea}. 

The Friedmann-Lemaître-Robertson-Walker (FLRW) metric corresponds to the most general spacetime exhibiting spatial homogeneity and isotropy. In $4$ dimensions, the metric can be written as 
\begin{equation}
 ds^2 = -dt^2 + a(t)^2d\Omega_k^2\ ,
\end{equation}
where $a(t)$ is called the scale factor and $d\Omega_k^2=\frac{d\rho^2}{1-k\rho^2}+\rho^2(d\theta^2+\sin^2\theta d\phi^2)$ is the $3$-dimensional spatial metric of a constant curvature space. The spatial curvature $k$ can be positive, negative, or zero, and without loss of generality we can set its possible values to be $k=+1,0,-1$ corresponding to a unit sphere, Euclidean space, or unit hyperbolic space, respectively. The FLRW metric describes an expanding (contracting) spacetime provided that $a(t)$ is a monotonically increasing (decreasing) function. It is largely used in cosmology due to the observation that our universe is homogeneous and isotropic (with $k=0$) in cosmological scales \cite{Dodelson:2003ft}. 

The FLRW spacetime also provides a good prototype to approach the nonequilibrium dynamics of an expanding system. It is well known, for instance, that even a locally static fluid flow characterized by the $4$-velocity $u^\mu=(1,0,0,0)$ in the FLRW metric has a nonvanishing expansion rate $\nabla_\mu u^\mu=3H(t)$ ($H(t)\equiv\frac{\dot{a}(t)}{a(t)}$ is the so-called Hubble parameter) due to the dynamical nature of the geometry itself. In spite of the time dependence, the high degree of spatial symmetry of the metric may render certain problems technically feasible. An interesting example is \cite{Bazow:2015dha}, where an analytical solution to the general relativistic Boltzmann equation in FLRW has been found describing the dynamics of an expanding massless gas with constant cross section.

In the present work, we use an expanding FLRW spacetime as the background arena in which to study a strongly coupled field theory plasma. In particular, we shall focus on a conformal field theory (CFT) with a holographic dual and resort to gauge/gravity duality methods in order to extract information about the expanding CFT plasma. An advantage of the holographic approach is that non-equilibrium quantities of the expanding plasma such as the 
entropy density can be naturally associated with the 
gravitational entropy of the apparent horizon on the gravity side \cite{Figueras:2009iu}. The crucial step here involves setting the Anti de Sitter (AdS) conformal boundary to take a FLRW form instead of the commonly used static boundaries. This is in principle possible since the boundary metric belongs to a conformal class, and one can switch between members of this class by appropriate bulk diffeomorphisms.

Previous efforts in this direction have been made in \cite{Apostolopoulos:2008ru} (see also \cite{Lamprou:2011sa,Tetradis:2009bk,Nojiri:2002hz}) using Fefferman-Graham coordinates, but were restricted basically to pure Einstein gravity in the bulk. We propose here a different foliation of AdS$_5$ black hole spacetimes using Eddington-Finkelstein-like coordinates such that the asymptotic boundary becomes the $4$-dimensional FLRW spacetime. The holographic dual picture is, therefore, that of a thermalized CFT on an expanding background, even though the bulk solution is not truly dynamical. The procedure is simple and applies equally well to a variety of AdS black holes, supported by external fields or not, leading to the same results of \cite{Apostolopoulos:2008ru} when applied to the AdS-Schwarzschild solution. 

The paper is organized as follows. In Section \ref{sec:AdSFLRWgeneral} we introduce our FLRW foliation of generic AdS black holes, discuss the entropy production by the dual expanding plasma and set the ground to discuss holographic renormalization and one-point functions in the sequence. In Section \ref{sec:AdSFLRWexamples} we give three explicit examples, namely for the AdS-Schwarzschild, AdS-Gauss-Bonnet, and AdS-Reissner-Nordström solutions, calculate the corresponding renormalized one-point functions of the dual CFT and discuss the conformal anomalies. Section \ref{sec:AdSFLRWconclusions} contains the closing remarks.

\section{AdS black holes with a FLRW boundary}\label{sec:AdSFLRWgeneral}

We begin with a generic asymptotically AdS$_5$ black hole written in the usual form
\begin{equation}\label{eq:BHmetricusual}
ds^2 = -f(r)dt^2+f(r)^{-1}dr^2+\Sigma(r)^2d\Omega_k^2\ ,
\end{equation}
where $d\Omega_k^2$ denotes the metric of the horizon, corresponding to a spherical, planar, or hyperbolic horizon for $k=+1,0,-1$ respectively. The blackening factor $f(r)$ and the function $\Sigma(r)$ are left completely general with the only assumption that $f(r)\sim\frac{r^2}{L^2}$ and $\Sigma(r)\sim\frac{r}{L}$ for large $r$, as required in order to have AdS asymptotics with curvature radius $L$.\footnote{Of course the zero temperature cases of Poincaré and global AdS are also included in this class and, incidentally, will be included in our analysis. However, we shall ignore them since we are interested only in CFTs at finite $T$.} The event horizon $r_h$ is defined by (the largest root of) $f(r_h)=0$ and the corresponding Hawking temperature is $T=|f'(r_h)|/4\pi$. There may eventually be matter fields supporting the geometry, but for our purposes at this section they will play no role.

The first step involves going to the so-called ingoing Eddington-Finkelstein coordinates ($v,r$). This is done by trading the time coordinate $t$ to a new coordinate $v$ adapted to ingoing null geodesics, which is defined by $dv=dt+f(r)^{-1}dr$. The metric then reads
\begin{equation}\label{eq:BHmetricEF}
ds^2 = 2dvdr - f(r)dv^2 + \Sigma(r)^2d\Omega_k^2\ .
\end{equation}
Notice that the large $r$ behavior of the metric is $ds^2 \sim 2dvdr + \frac{r^2}{L^2}[- dv^2 + d\Omega_k^2]$, from where it is clear that the $4$-dimensional conformal boundary at $r=\infty$ (where the dual CFT lives) is the Einstein static universe $\mathbb{R}\times\Sigma_k$ with metric\footnote{The time coordinates $v$ and $t$ coincide at fixed-$r$ surfaces.}
\begin{equation}\label{eq:ESUk}
 ds_0^2 = g_{\mu\nu}^{(0)}dx^\mu dx^\nu = -dt^2+d\Omega_k^2\ .
\end{equation}
When $k=0$ this is just the $4$-dimensional Minkowski spacetime, which is by far the most studied one in holographic applications since most field theories of physical interest live in flat spacetime. In spite of that, for the sake of completeness we shall keep the spatial curvature $k$ arbitrary in the sequence. 

Before proceeding it is instructive to recall a simple reason why the AdS conformal boundary indeed goes well with the intuitive notion of a boundary. This can be seen by calculating the time interval $\Delta v(r_0)$ spent by an outgoing light ray to travel radially from $r_0>r_h$ to the boundary at $r=\infty$ and back to $r_0$. It follows immediately from the definition of outgoing null geodesics in \eq{BHmetricEF}, $2dr-f(r)dv=0$, that
\begin{equation}
 \Delta v(r_0) = 4 \int_{r_0}^\infty\ \frac{dr'}{f(r')} < \infty\ ,
\end{equation}
which is obviously finite since there are no poles in the denominator for $r_0>r_h$ and the integrand vanishes at large $r'$.

In the following we shall introduce a different foliation of the black hole spacetime \eq{BHmetricEF} in such a way that the corresponding conformal boundary takes the form of a FLRW spacetime, namely
\begin{equation}\label{eq:FLRWk}
 ds_0^2 = g_{\mu\nu}^{(0)}dx^\mu dx^\nu = -dt^2+a(t)^2 d\Omega_k^2\ .
\end{equation}

In order to achieve that one needs two further coordinate transformations. We first define a new time coordinate $V$ with respect to which the old $v$ is a \lq\lq conformal time\rq\rq~with scale factor $a(V)$, i.e., $dv=\frac{dV}{a(V)}$, where $a(V)$ is assumed to be everywhere continuous and non-vanishing. 
Finally, we introduce a new radial coordinate $R$ defined as $R=\frac{r}{a(V)}$. After plugging $dv=\frac{dV}{a}$ and $dr=a(V)dR+R\dot{a}(V)dV$ the metric \eq{BHmetricEF} becomes
\begin{eqnarray}\label{eq:BHmetricFLRW}
ds^2 &=& 2\frac{dV}{a}(adR+R\dot{a}dV)-f(Ra)\frac{dV^2}{a^2}+\Sigma(Ra)^2d\Omega_k^2\nonumber\\
&=& 2dVdR-\left[\frac{f(Ra)}{a^2}-2R\frac{\dot{a}}{a}\right]dV^2+\Sigma(Ra)^2d\Omega_k^2
\end{eqnarray}
This form is the one we are interested in this work. Note that the metric is still expressed in Eddington-Finkelstein-like coordinates (in the sense that $V$ is still adapted to null geodesics), but now it carries an artificial time dependence reminiscent of the transformation from $v$ to $V$. At large $R$ we have $f(Ra)\sim\frac{(Ra)^2}{L^2}$ and $\Sigma(Ra)\sim\frac{Ra}{L}$ due to our assumption of AdS asymptotics, and, therefore, 
$$ds^2\sim 2dVdR+\frac{R^2}{L^2}[-dV^2+a(V)^2d\Omega_k^2]\ .$$ 
As a result, the new conformal boundary at $R=\infty$ has precisely the desired FLRW form \eq{FLRWk} with spatial curvature $k$ (the time coordinate is now called $V$). We shall refer to this as the \emph{cosmological boundary} just to remind that this is not the same as the commonly used AdS boundary at $r=\infty$. 

Actually we shall pause for a moment here to argue that the cosmological boundary introduced above is indeed also compatible with the notion of a boundary. This is done by asking the same question asked previously for the AdS boundary, namely whether the time interval $\Delta V(R_0)$ spent by an outgoing light ray to go radially from $R_0$ to $\infty$ and back to $R_0$ in the metric \eq{BHmetricFLRW} is finite or not. The outgoing null geodesics in this case satisfy
\begin{equation}
 \frac{dR}{dV}=\frac{1}{2}\left[\frac{f(Ra)}{a^2}-2R\frac{\dot{a}}{a}\right] = \frac{1}{2}\left[\frac{R^2}{L^2}-2R\frac{\dot{a}}{a}+\cdots\right]\ .
\end{equation}
For simplicity we focus on the case of pure AdS space ($f(r)=\frac{r^2}{L^2}$), without loss of generality since this corresponds to the asymptotic structure of any AdS black hole. In this case the ellipsis in the previous expression is not present and it can be exactly integrated to yield 
\begin{equation}
 R(V) = \frac{R_0a_0}{a(V)\left[1-\frac{R_0a_0}{2L^2}\int_{V_0}^V\ \frac{dV'}{a(V')}\right]}\ ,
\end{equation}
where we have introduced $R_0\equiv R(V_0)$ and $a_0\equiv a(V_0)$. The time $V_\infty$ corresponding to reaching the cosmological boundary $R=\infty$ is implicitly defined by
\begin{equation}
 \frac{R_0a_0}{2L^2}\int_{V_0}^{V_\infty}\ \frac{dV'}{a(V')}=1\ .
\end{equation}
A straightforward consequence of our assumption that the scale factor $a(V)$ is a continuous and everywhere non-vanishing function is that $V_\infty$ must be finite (although an explicit expression for it cannot be obtained without specifying the form of the scale factor). As a result, the time interval $\Delta V(R_0)=2(V_\infty-V_0)$ is guaranteed to be finite and, therefore, $R=\infty$ also provides a sensible notion of asymptotic boundary. 


To summarize, we have introduced a different type of foliation for $5$-dimensional AdS black hole spacetimes of the form \eq{BHmetricusual} where the $4$-dimensional slices asymptotically approach the FLRW metric. This is very similar to the work done in \cite{Apostolopoulos:2008ru}. It should be stressed, however, that our procedure is astonishingly simpler and, in particular, our metric \eq{BHmetricFLRW} applies equally well for any AdS black hole (characterized by the functions $f,\Sigma$ and eventually matter fields\footnote{Of course matter fields when present must be changed according to the same coordinate transformations above to take a different $(V,R)$-dependent configuration that supports the FLRW-foliated metric \eq{BHmetricFLRW}. This is the case, e.g., for the AdS-Reissner-Nordström charged black hole (see Section \ref{sec:AdSFLRWexamples}).}), as we shall illustrate in the sequence, while the method of \cite{Apostolopoulos:2008ru} is hardly applicable beyond the simplest case of the AdS-Schwarzschild solution. 

\subsection{Entropy production}\label{sec:EntropyProduction}

We begin our analysis by finding the location of the apparent horizon in our FLRW-foliated black hole metric \eq{BHmetricFLRW}. 
The apparent horizon is formally defined as the outermost trapped surface, i.e., the closed null hypersurface on which all radially outgoing null geodesics have vanishing expansion (see e.g. \cite{booth:2005qc}). For a generic $5$-dimensional metric of the form $$ds^2=2dVdR-\alpha(V,R)dV^2+\beta(V,R)^2d\Omega_k^2$$ the expansion along outgoing null rays is given by $\theta_\text{out}\equiv\left(\partial_V+\frac{\alpha}{2}\partial_{R}\right)\ln\beta^3$ and the apparent horizon hence corresponds to the location $R_h(V)$ for which $\theta_\text{out} = 0$. For the case of interest \eq{BHmetricFLRW}, with $\alpha(V,R)=\frac{f(Ra)}{a^2}-2R\frac{\dot{a}}{a}$ and $\beta(V,R)=\Sigma(Ra)$, the result is
\begin{equation}\label{eq:apparenthorizonFLRWdef}
 \left[\partial_V\Sigma + \biggl(\frac{f(Ra)}{2a^2}-R\frac{\dot{a}}{a}\biggr)\partial_R\Sigma\right]\bigg|_{R=R_h} = 0\ .
\end{equation}
However, since $\Sigma$ only depends on $(V,R)$ through the combination $Ra(V)$, one can write $\partial_V\Sigma(Ra)=R\dot{a}\Sigma'(Ra)$, $\partial_R\Sigma(Ra)=a\Sigma'(Ra)$ (here a prime denotes the derivative with respect to the argument $Ra$) and, therefore, the definition of $R_h$ reduces to
\begin{equation}
 f(R_ha)\Sigma'(R_ha) = 0\ .
\end{equation}
For most black hole solutions of interest, which have $\Sigma(r)=\frac{r}{L}$ (see next section), the equation above gives simply $f(R_ha)=0$ or, equivalently, $R_h(V)=\frac{r_h}{a(V)}$, where the constant $r_h$ denotes the black hole event horizon in the standard coordinate system \eq{BHmetricEF} (i.e., $f(r_h)=0$). Nevertheless, if $\Sigma(r)$ has subleading contributions in $r$ of any kind such that $\Sigma'(r)$ is not constant, there may appear an additional horizon corresponding to $\Sigma'(R_ha)=0$.

Having determined its location, we now follow \cite{Figueras:2009iu} and associate the non-equilibrium entropy density $s$ of the expanding plasma living at the cosmological boundary with the Bekenstein-Hawking entropy of the apparent horizon, namely
\begin{equation}
 s = \frac{\Sigma(R_ha)^3}{4G_5}\ .
\end{equation}
From this it follows that, if $R_h=\frac{r_h}{a}$ is the only apparent horizon, then clearly
\begin{equation}
\frac{ds}{dV} = 0\ ,
\end{equation}
i.e., there is no entropy production by the plasma during the dynamical process. This matches the expectation from the hydrodynamics of conformally invariant fluids, for which there is no entropy production at all orders in the hydrodynamic expansion (see \cite{Buchel:2016cbj}). However, if the bulk solution admits another apparent horizon corresponding to the root of $\Sigma'(R_ha)$, then there may be a nonzero entropy production by the plasma since the combination $R_ha$ on which $s$ depends will not necessarily be constant anymore. This is the case, for instance, for the $\mathcal{N}=2^*$ plasma studied in \cite{Buchel:2016cbj}.

Similarly, one can associate to the expanding plasma the local temperature 
\begin{equation}\label{eq:LocalT}
 T(V) = \frac{T_H}{a(V)}\ ,
\end{equation}
where $T_H$ is the temperature of the corresponding static plasma (i.e., the Hawking temperature of the black hole). As argued in \cite{Apostolopoulos:2008ru}, this follows from the fact that the FLRW metric \eq{FLRWk} and the static boundary metric \eq{ESUk} are connected by a Weyl rescaling, i.e., $ds^2_{\text{FLRW}}=a(\eta)^2ds_0^2$ where $\eta\equiv\int{\tfrac{dt}{a}}$ is the conformal time. As a result, the local temperature of the plasma in FLRW and the equilibrium temperature $T_H$ of the static plasma must be linked by a rescaling. Since the Euclidean proper time period in FLRW 
scales as $a$ according to the formula above, the temperature of the expanding plasma, being inversely related to the period, must scale as $a^{-1}$ with respect to $T_H$. Another way to see that is to recall that our new slicing does not change the physical content of the bulk solution, i.e., we still have the same static AdS black hole in thermal equilibrium with its Hawking radiation at temperature $T_H$. The difference now is that we have a new notion of boundary ($R=\infty$) that expands in time according to the scale factor $a(V)$, and a comoving observer sitting in there will experience a temperature appropriately corrected by $a$ that corresponds precisely to \eq{LocalT}.

\subsection{One-point functions}
We now follow the spirit of \cite{Apostolopoulos:2008ru} and, by assuming that
\emph{i)} the cosmological boundary is holographic;
\emph{ii)} the standard holographic renormalization procedure can be carried out in the same way in there as in the usual AdS boundary, 
we proceed to calculate the one-point functions for the dual CFT operators living in the cosmological boundary, i.e., for CFTs in FLRW spacetime.

The first step involved is to find the Fefferman-Graham (FG) expansion of the bulk metric (and eventually matter fields) near the cosmological boundary, since knowledge of the FG coefficients determines the CFT correlators via the holographic dictionary. Namely, we need to put the metric \eq{BHmetricFLRW} in the FG form
\begin{equation}\label{eq:FGmetric}
ds^2 = \frac{L^2}{z^2}\left[dz^2 + g_{\mu\nu}(z,x)dx^\mu dx^\nu\right]
\end{equation}
(here $z\sim L^2/R$ such that the cosmological boundary is, in these coordinates, at $z=0$) and find the first few coefficients of the near boundary expansion of $g_{\mu\nu}$, 
\begin{eqnarray}\label{eq:FGmetricasymptotic}
 g_{\mu\nu}(z,x) &=& g_{\mu\nu}^{(0)}(x) + z^2g_{\mu\nu}^{(2)}(x) +z^4\bigl(g_{\mu\nu}^{(4)}(x)+h_{\mu\nu}^{(4)}(x)\log z\bigr)+ \cdots\ ,
\end{eqnarray}
where the leading one, $g_{\mu\nu}^{(0)}(x)$, is the FLRW metric \eq{FLRWk}, and the subleading ones are determined by the bulk equations of motion.
A practical way to achieve that is to write generic coordinate transformations from ($V,R$) to FG coordinates ($x^0\equiv\tau,z$), i.e., $V=V(\tau,z)$ and $R=R(\tau,z)$, and then get the transformation equations by comparing our metric \eq{BHmetricFLRW} with \eq{FGmetric}. This leads to the following set of equations 
\begin{eqnarray}\label{eq:VRtoFGeqs}
2\partial_z R\partial_z V-\alpha (\partial_z V)^2 &=& \frac{L^2}{z^2}\nonumber\\
\partial_z V\partial_\tau{R}+\partial_\tau{V}\partial_z R-\alpha \partial_z V\partial_\tau{V} &=& 0\ ,
\end{eqnarray}
which determine the precise form of the transformations, together with the FG metric components expressed in terms of $V$ and $R$, which can be massaged to take the simple form
\begin{equation}\label{eq:FGmetriccoeffsVR}
 g_{\tau\tau}=-\frac{(\partial_\tau{V})^2}{(\partial_z V)^2}\qquad g_{ij}dx^idx^j=\frac{z^2}{L^2}\Sigma^2d\Omega_k^2\qquad g_{\tau i}=0\ .
\end{equation}
The near-boundary solution to the transformation equations \eq{VRtoFGeqs} can be easily obtained to any desired order with a power series ansatz of the form
\begin{equation}\label{eq:VRtoFGsolgeneric}
V(\tau,z) = \sum_{n=0}V_n(\tau)z^n\qquad R(\tau,z) = \sum_{n=0}R_n(\tau)z^{n-1} 
\end{equation}
with $V_0(\tau)\equiv\tau$ (such that $V$ and $\tau$ coincide at the boundary) and $R_0(\tau)\equiv L^2$ (such that $R=\frac{L^2}{z}+\cdots$). Once this solution is found, by plugging it back into \eq{FGmetriccoeffsVR} and expanding for small $z$ yields the desired FG asymptotic expansion \eq{FGmetricasymptotic}. One is then ready to obtain the corresponding one-point functions of the dual CFT living on the cosmological boundary using standard holographic renormalization.

So far the analysis has been quite general. We shall now illustrate the procedure by particularizing the functions $f(r),\Sigma(r)$ to a few black holes of physical interest.

\section{Examples}\label{sec:AdSFLRWexamples}

\subsection{AdS-Schwarzschild black hole}

The AdS-Schwarzschild black hole is an exact static solution to pure Einstein gravity with a negative cosmological constant $\Lambda=-12/L^2$ in the bulk, namely
\begin{equation}
 S = \frac{1}{16\pi G_5}\int\ d^5x\sqrt{-g}\left[R+\frac{12}{L^2}\right].
\end{equation}
The solution with horizon curvature $k$ corresponds to a metric of the form \eq{BHmetricusual} with $f(r)$ and $\Sigma(r)$ given by
\begin{equation}
 f(r) = \frac{r^2}{L^2}\left(1+\frac{kL^2}{r^2}-\frac{M}{r^4}\right)\qquad \Sigma(r)=\frac{r}{L}\ ,
\end{equation}
where the mass $M$ is related to the event horizon radius $r_h$ according to 
$M = r_h^4\left(1+\frac{kL^2}{r_h^2}\right)$. 
Its corresponding Hawking temperature is readily found to be
\begin{equation}\label{eq:HawkingTSAdS5k}
 T_H = \frac{kL^2+2r_h^2}{2\pi L^2r_h}\ .
\end{equation}

The explicit form of the foliation \eq{BHmetricFLRW} for the AdS-Schwarzschild black hole reads
\begin{eqnarray}\label{eq:SAdSmetricFLRW}
 ds^2 &=& 2dVdR-\left[\frac{R^2}{L^2}\left(1+\frac{kL^2}{R^2a^2}-\frac{M}{R^4a^4}\right)-2R\frac{\dot{a}}{a}\right]dV^2+\frac{R^2a^2}{L^2}d\Omega_k^2\ .
\end{eqnarray}
If the standard holographic dictionary extrapolates to the cosmological boundary $R=\infty$, this metric would be the holographic dual of a $\mathcal{N}=4$ SYM plasma in the FLRW metric \eq{FLRWk} with spatial curvature $k$. As discussed in Section \ref{sec:EntropyProduction}, one can associate to this nonequilibrium plasma the local temperature \eq{LocalT}, namely
\begin{equation}\label{eq:localHawkingTSAdS5k}
 T(V)=\frac{kL^2+2r_h^2}{2\pi L^2r_ha}\ .
\end{equation}
In the following for the sake of simplicity we take $L=1$.

It is worth mentioning that the metric above in the planar case ($k=0$) has been previously used by the authors of \cite{Buchel:2016cbj} as the starting point to study the $\mathcal{N}=2^*$ plasma close to conformality in a FLRW spacetime.\footnote{We emphasize that, although not obvious, this is nothing but AdS-Schwarzschild expressed in unusual coordinates.} 


The transformation \eq{VRtoFGsolgeneric} from our ($V,R$) coordinates to the Fefferman-Graham system ($\tau,z$) is given asymptotically by
\begin{eqnarray}\label{eq:VRtoFGsolSAdS}
V(\tau,z) &=& \tau -z +\frac{-2a\ddot{a}+\dot{a}^2+k}{12a^2}z^3 +\frac{a^2 \dddot{a}+\dot{a}^3+\dot{a}\left(k-2 a\ddot{a}\right)}{24a^3}z^4 \notag\\ &&+\frac{3\dot{a}^4-2{a}^{(4)}a^3-3\left(k^2+6M\right)+2a\ddot{a}\left(5k-\dot{a}^2\right)+a^2\left(6\dddot{a}\dot{a}-8\ddot{a}^2\right)}{240a^4}z^5 + \cdots\notag\\
R(\tau,z) &=& \frac{1}{z} + \frac{\dot{a}}{a} - \frac{2a\ddot{a}-3\dot{a}^2+k}{4a^2}z + \frac{a^2\dddot{a}+4\dot{a}^3-\dot{a}\left(5a\ddot{a}+2k\right)}{6a^3}z^2 \notag\\
&&+ \frac{13\dot{a}^4-a^3{a}^{(4)}-11k\dot{a}^2+a\ddot{a}\left(5k-21\dot{a}^2\right)+a^2\left(2\ddot{a}^2+7\dddot{a}\dot{a}\right)+3M}{24a^4}z^3 + \cdots\notag\\
\end{eqnarray}
with $a$ and its derivatives now viewed as functions of $\tau$. From \eq{FGmetriccoeffsVR} it then follows that the Fefferman-Graham expansion \eq{FGmetricasymptotic} of the metric in this case has the following non-null coefficients
\begin{align}\label{eq:FGcoeffsSAdS}
&g_{\tau\tau}^{(0)} = -1\notag\\
&g_{\tau\tau}^{(2)} = -\frac{\dot{a}^2-2a\ddot{a}+k}{2a^2}\notag\\
&g_{\tau\tau}^{(4)} = -\frac{\dot{a}^4+4a^2 \ddot{a}^2+2\dot{a}^2(k-2a\ddot{a})-4ka\ddot{a}+k^2-12M}{16a^4}\notag\\
&g_{ij}^{(0)}dx^idx^j = a^2d\Omega_k^2\notag\\
&g_{ij}^{(2)}dx^idx^j = -\frac{\dot{a}^2+k}{2}d\Omega_k^2\notag\\
&g_{ij}^{(4)}dx^idx^j = \frac{2k\dot{a}^2+\dot{a}^4+k^2+4 M}{16a^2}d\Omega_k^2\ .
\end{align}

The holographic renormalization for pure Einstein gravity in the bulk has been done in \cite{deHaro:2000vlm}, to which we refer the reader for details. The resulting expression for the renormalized energy-momentum tensor of the dual CFT living on the boundary with metric $g^{(0)}$ is generically given by
\begin{eqnarray}\label{eq:TmunuCFTeinstein}
\langle T_{\mu\nu}\rangle &=& \frac{1}{ 4\pi G_5}\left\{g^{(4)}_{\mu\nu}-\frac{1}{2}g^{(2)\sigma}_{\ \mu}g^{(2)}_{\sigma\nu}
+\frac{1}{4}\bigl(g^{(2)\sigma}_{\ \sigma}\bigr) g^{(2)}_{\mu\nu}-\frac{1}{8}\bigl[\bigl(g^{(2)\sigma}_{\ \sigma}\bigr)^2-g^{(2)}_{\sigma\rho}g^{(2)\rho\sigma}
\bigr]g^{(0)}_{\mu\nu}\right\}\ ,
\end{eqnarray}
where indices are to be raised and lowered with the boundary metric $g^{(0)}$. 
In our case, with the FG coefficients \eq{FGcoeffsSAdS}, this yields the following energy density $\mathcal{E}\equiv \langle T_{\tau\tau}\rangle$ and pressure $\mathcal{P}\equiv \langle T_{\ i}^{i}\rangle$ (no summation over $i$ implied) for the $\mathcal{N}=4$ SYM plasma 
\begin{eqnarray}\label{eq:CFTenergypressureSAdS}
\mathcal{E} &=& \frac{3(\dot{a}^2+k)^2+12M}{64\pi G_5 a^4} \nonumber\\
\mathcal{P} &=& \frac{(\dot{a}^2+k)^2+4M-4a\ddot{a}(\dot{a}^2+k)}{64\pi G_5 a^4}\ ,
\end{eqnarray}
in perfect agreement with the results of \cite{Apostolopoulos:2008ru}. These expressions can be cast entirely in $4$-dimensional CFT language by expressing the mass parameter $M = r_h^4\left(1+\frac{kL^2}{r_h^2}\right)$ in terms of the local temperature $T$ of the plasma using \eq{localHawkingTSAdS5k} and the $5$-dimensional Newton constant $G_5$ in terms of the number of colors $N_c$ via the standard AdS$_5$/CFT$_4$ relation $G_5=\frac{\pi L^3}{2N_c^2}$. For instance, in the $k=0$ case, with $M=(\pi aT)^4$ we obtain
\begin{eqnarray}\label{eq:CFTenergypressureSAdSin4dlanguagek=0}
\mathcal{E} &=& \frac{3N_c^2T^4}{8}+\frac{3N_c^2}{32\pi^2}\frac{\dot{a}^4}{a^4}\nonumber\\
\mathcal{P} &=& \frac{\mathcal{E}}{3}-\frac{N_c^2}{8\pi^2}\frac{\ddot{a}\dot{a}^2}{a^3}\ .
\end{eqnarray}

It interesting to note that when $a(V)\equiv1$ (where $R=\infty$ becomes the usual AdS boundary $r=\infty$) we get the expected conformal plasma in $\mathbb{R}\times\Sigma_k$ with $\mathcal{E}=3\mathcal{P}$, while the presence of a non-constant scale factor breaks the conformal invariance leading to a conformal anomaly given by
\begin{equation}\label{eq:CFTanomalySAdS}
 \langle T_{\ \mu}^{\mu}\rangle = 3\mathcal{P}-\mathcal{E} = -\frac{3\ddot{a}(\dot{a}^2+k)}{16\pi G_5 a^3}\ . 
\end{equation}
The anomaly has a clearly geometric nature due exclusively to the nontrivial rate of cosmological expansion. For an expanding plasma ($\ddot{a}>0$) in flat space or in a sphere this quantity is strictly negative. 

\subsection{AdS-Gauss-Bonnet black hole}

We start by reviewing the Einstein-Gauss-Bonnet action in $5$ dimensions. It consists in one of the simplest generalizations of Einstein gravity built from higher derivative terms in the action that still yield second order equations of motion for the metric.\footnote{In fact, the Einstein-Gauss-Bonnet action is just a very special case of the so-called \emph{Lovelock gravity}, which is the most general metric theory of gravity giving rise to second order equations of motion (see \cite{Padmanabhan:2013xyr} for a review).} With the inclusion of a negative cosmological constant $\Lambda\equiv-12/L^2$, the action is
\begin{eqnarray}\label{eq:EGBaction}
S &=& \frac{1}{16\pi G_5}\int d^5x\sqrt{-g}\Big[R+\frac{12}{L^2}+\frac{L^2}{2}\gb(R_{abcd}R^{abcd}-4R_{ab}R^{ab}+R^2)\Big], 
\end{eqnarray}
where $\gb$ is the Gauss-Bonnet parameter. It is still unclear at the moment whether a higher curvature correction of the Gauss-Bonnet type \eq{EGBaction} can be obtained from a top-down string theory construction: the leading $\alpha'$ corrections to the action of Type IIB supergravity, corresponding to finite 't Hooft coupling corrections to the dual $\mathcal{N}=4$ SYM theory, are known to take the form of more complicated higher curvature terms schematically of the form $\alpha'^3R^3$ \cite{Grisaru:1986vi}. Nevertheless, the general belief is that it may provide qualitative information into properties shared by generic higher curvature terms, with the practical advantages of being tractable and having a number of exact solutions available in the literature.

The action \eq{EGBaction} has been extensively studied in the context of holography. Interestingly, the presence of the extra Gauss-Bonnet coupling $\gb$ in the bulk allows for a holographic dual CFT with two distinct central charges $c\ne b$ \cite{Nojiri:1999mh,Blau:1999vz}.\footnote{We denote here the second central charge by $b$ instead of the commonly used $a$ in order to avoid confusion with the scale factor $a(V)$ appearing throughout the paper.} Namely, the central charges, defined in the standard way via the conformal anomaly as
\begin{equation}\label{eq:GBcentralchargesdef}
 \langle T_{\ \mu}^{\mu}\rangle = \frac{1}{16\pi^2}(cW-bE)\ ,
\end{equation}
are related to $\gb$ and the other gravitational parameters via \cite{deBoer:2011wk}
\begin{eqnarray}\label{eq:GBcentralchargesdef2}
c &=& \frac{\pi L_\text{AdS}^3}{8G_5}\sqrt{1-4\gb} \notag\\
b &=& \frac{\pi L_\text{AdS}^3}{8G_5}\bigl(-2+3\sqrt{1-4\gb}\bigr)\ .
\end{eqnarray}
The AdS radius appearing above depends on $\gb$ 
(see below for details), while the quantities 
$W\equiv W_{\mu\nu\rho\sigma}W^{\mu\nu\rho\sigma}$ and $E\equiv\mathcal{R}_{\mu\nu\sigma\rho}\mathcal{R}^{\mu\nu\sigma\rho}-4\mathcal{R}_{\mu\nu}\mathcal{R}^{\mu\nu}+\mathcal{R}^2$ are respectively the squared Weyl tensor and the Euler density associated with the $4$-dimensional metric where the CFT lives. In the Einstein gravity limit $\gb=0$ the two central charges collapse to a single one $c=b\sim N_c^2$ and the $SU(N_c)$ $\mathcal{N}=4$ SYM theory is recovered consistently.

The AdS-Gauss-Bonnet black hole with horizon curvature $k$ is an exact static spherically symmetric solution to the equations of motion of \eq{EGBaction}, first obtained in \cite{Cai:2001dz}. The metric has the standard black hole form \eq{BHmetricEF} with
\begin{eqnarray}
 f(r) &=& k + \frac{r^2}{2L^2\gb} \left[1-\sqrt{1-4\gb\left(1-\frac{ML^2}{r^4}\right)}\ \right]\notag\\
 \Sigma(r)&=&\frac{r}{L}\ ,
\end{eqnarray}
where $M$ is a parameter related to the black hole mass that can be conveniently expressed in terms of the event horizon location $r_h$ as $$M\equiv r_h^4\left(\frac{1}{L^2} + \frac{k}{r_h^2}+\gb \frac{L^2k^2}{r_h^4}\right)\ .$$ 
The Hawking temperature associated to this solution reads
\begin{equation}\label{eq:localHawkingTGBAdS5k}
T_H = \frac{r_h(2r_h^2+kL^2)}{2\pi L^2(r_h^2+2kL^2\gb)}\ .
\end{equation}

It is worth noticing that the AdS-Gauss-Bonnet is an asymptotically AdS black hole, i.e., $f(r)\sim\frac{r^2}{L_{_\text{AdS}}^2}$ for large $r$. However, the AdS curvature radius is shifted from the usual $L$ to an effective radius $L_\text{AdS}$ due to the presence of $\gb$, namely $L_\text{AdS}^2\equiv\frac{L^2}{2}(1+\sqrt{1-4\gb})$. In particular, when the standard choice of units $L=1$ is made (which corresponds to making the cosmological constant $\Lambda=-12$) it should be kept in mind that the resulting AdS radius appearing in the metric is not unity.

Our FLRW foliation \eq{BHmetricFLRW} of the AdS-Gauss-Bonnet black hole metric takes the form
\begin{align}
ds^2 = 2dVdR+\frac{R^2a^2}{L_\text{eff}^2}d\Omega_k^2-\left\{\frac{k}{a^2}+\frac{R^2}{2L^2\gb}\bigg[1-\sqrt{1-4\gb\Big(1-\frac{ML^2}{R^4a^4}\Big)}\bigg]-2R\frac{\dot{a}}{a}\right\}dV^2
\end{align}
where the spatial coordinates were conveniently redefined by appropriate factors so as to make $\frac{d\Omega_k^2}{L^2}\rightarrow\frac{d\Omega_k^2}{L_\text{AdS}^2}$ and, hence, have a canonically normalized FLRW boundary of the form \eq{FLRWk}. Just as in the AdS-Schwarzschild case (see previous section), the holographic dual expanding CFT plasma living at $R=\infty$ can be associated the local temperature $T(V)=\tfrac{T_H}{a(V)}$.

From now on we shall take $L=1$ and treat the Gauss-Bonnet parameter as small, working always to linear order in $\gb$ for simplicity (hence all the formulas containing $\gb$ below are valid up to $\mathcal{O}(\gb^2)$ corrections, although we choose not to unnecessarily repeat this symbol in each and every expression). The transformation from ($V,R$) to the Fefferman-Graham coordinates ($\tau,z$) can be obtained precisely in the same way as before (the expressions are too cumbersome to be shown here, though),
from where we get the following FG metric coefficients 
\begin{align}\label{eq:FGcoeffsGB}
g_{\tau\tau}^{(0)} &= -1\notag\\
g_{\tau\tau}^{(2)} &= -\frac{1}{2a^2}(1-\gb)\left(-2a\ddot{a}+\dot{a}^2+k\right) \notag\\
g_{\tau\tau}^{(4)} &= -\frac{1}{16a^4}\bigl[(1-2\gb)\bigl(4a\ddot{a}(a\ddot{a}-\dot{a}^2-k)+2k\dot{a}^2+\dot{a}^4+k^2\bigr)-12(1+\gb)M\bigr]\notag\\
g_{ij}^{(0)}dx^idx^j &= a^2d\Omega_k^2\notag\\
g_{ij}^{(2)}dx^idx^j &= -\frac{1}{2}(1-\gb) \left(\dot{a}^2+k\right)d\Omega_k^2\notag\\
g_{ij}^{(4)}dx^idx^j &= \frac{1}{16a^2}\bigl[(1-2\gb)(2k\dot{a}^2+\dot{a}^4+k^2)+4(1+\gb)M\bigr]d\Omega_k^2\ .
\end{align}

The holographic renormalization of the Einstein-Gauss-Bonnet action \eq{EGBaction} has been carried out in detail in \cite{Liu:2008zf,Astefanesei:2008wz,Jahnke:2014vwa} to linear order in $\gb$ (see also \cite{Brihaye:2008xu} for arbitrary $\gb$).\footnote{We follow here the same conventions of \cite{Jahnke:2014vwa}. In particular, the Gauss-Bonnet parameter $\gb$ used here differs from the $\alpha$ used in \cite{Liu:2008zf}, namely $\alpha\equiv\frac{L^2}{2}\gb$.} The final expression for the boundary energy-momentum tensor can be expressed in terms of the Fefferman-Graham coefficients as
\begin{equation}
\langle T_{\mu\nu}\rangle =  \langle T_{\mu\nu}\rangle_{_\text{Einstein}} + \gb \langle T_{\mu\nu}\rangle_{_\text{GB}}\ ,
\end{equation}
where the $\gb=0$ contribution $\langle T_{\mu\nu}\rangle_{_\text{Einstein}}$ due to pure Einstein gravity is the same as in \eq{TmunuCFTeinstein}, while the first order Gauss-Bonnet correction $\langle T_{\mu\nu}\rangle_{_\text{GB}}$ reads
\begin{eqnarray}
 \langle T_{\mu\nu}\rangle_{_\text{GB}}  &=& \frac{1}{16\pi G_5}\bigl[
-4g^{(2)\sigma}_{\ \mu}g^{(2)}_{\nu\sigma} + 7g^{(2)\sigma}_{\ \sigma}g^{(2)}_{\mu\nu} - 6g^{(4)}_{\mu\nu} - g^{(2)}_{\sigma\rho} g^{(2)}{}^{\sigma\rho} g^{(0)}_{\mu\nu} - 2\bigl(g^{(2)\sigma}_{\ \sigma}\bigr)^2 g^{(0)}_{\mu\nu} \nonumber\\
&& + 6g^{(4)\sigma}_{\ \sigma} g^{(0)}_{\mu\nu} - 3h^{(4)}_{\mu\nu} + 3h^{(4)\sigma}_{\ \sigma} g^{(0)}_{\mu\nu} + \tfrac{13}{4} \mathcal{R}^{(0)} g^{(2)}_{\mu\nu} - 2\mathcal{R}^{(0)}g^{(2)\sigma}_{\ \sigma} g^{(0)}_{\mu\nu}\nonumber \\ 
&& + \tfrac{29}{2}{g^{(2)}}{}^{\sigma\rho}\mathcal{R}^{(0)}_{\mu\sigma\nu\rho} + 4g^{(2)\sigma}_{\ \sigma} \mathcal{R}^{(0)}_{\mu\nu} - \tfrac{53}{4}g^{(2)\sigma}_{\ \nu}\mathcal{R}^{(0)}_{\mu\sigma} - \tfrac{53}{4}g^{(2)\sigma}_{\ \mu}\mathcal{R}^{(0)}_{\nu\sigma} +  \tfrac{11}{4}{g^{(2)}}{}^{\sigma\rho}\mathcal{R}^{(0)}_{\sigma\rho} g^{(0)}_{\mu\nu}\nonumber \\ 
&& + \tfrac{37}{4}\nabla^{(0)}_{\nu}\nabla^{(0)}_{\mu}g^{(2)\sigma}_{\ \sigma} - \tfrac{37}{4}\nabla^{(0)}_{\mu}\nabla^{(0)}_{\sigma}g^{(2)\sigma}_{\ \nu} - \tfrac{37}{4}\nabla^{(0)}_{\nu}\nabla^{(0)}_{\sigma}g^{(2)\sigma}_{\ \mu} + \tfrac{5}{4}g^{(0)}_{\mu\nu} \nabla^{(0)}_{\rho}\nabla^{(0)}_{\sigma}{g^{(2)}}{}^{\sigma\rho}\nonumber \\ 
&& + \tfrac{37}{4}\Box^{(0)}g^{(2)}_{\mu\nu} - \tfrac{5}{4}g^{(0)}_{\mu\nu}\Box^{(0)}g^{(2)\sigma}_{\ \sigma}\bigr]\ .
\end{eqnarray}
In the above the covariant derivatives $\nabla^{(0)}$ as well as the curvatures $\mathcal{R}^{(0)}$ are to be calculated with the boundary metric $g^{(0)}_{\mu\nu}$. In our case, inserting the FG coefficients \eq{FGcoeffsGB} and expanding to linear order in $\gb$ results in the following energy density $\mathcal{E}\equiv \langle T_{\tau\tau}\rangle$ and pressure $\mathcal{P}\equiv \langle T_{\ i}^{i}\rangle$ of the dual CFT 
\begin{eqnarray}\label{eq:CFTenergypressureAdSGB}
\mathcal{E} &=& \frac{3(\dot{a}^2+k)^2+12M}{64\pi G_5 a^4}-\frac{3\bigl[15(k+\dot{a}^2)^2+4M-64a\ddot{a}(k+\dot{a}^2-a\ddot{a})\bigr]}{128\pi G_5 a^4}\gb \nonumber\\
\mathcal{P} &=& \frac{(\dot{a}^2+k)^2+4M-4a\ddot{a}(k+\dot{a}^2)}{64\pi G_5 a^4}-\frac{15(k+\dot{a}^2)^2+4M-4a\ddot{a}[31(k+\dot{a}^2)-16a\ddot{a}]}{128\pi G_5a^4}\gb\ .\notag\\
\end{eqnarray}
Notice, in particular, that there is a conformal anomaly given by 
\begin{eqnarray}\label{eq:CFTanomalyGBAdS}
 \langle T_{\ \mu}^{\mu}\rangle = 3\mathcal{P}-\mathcal{E} =  -\Bigl(1-\frac{15}{2}\gb\Bigr)\frac{3\ddot{a}(k+\dot{a}^2)}{16\pi G_5 a^3}  \ ,
\end{eqnarray}
in agreement with the generic structure \eq{GBcentralchargesdef}. 
Namely, the central charges $c$ and $b$ of \eq{GBcentralchargesdef2} when linearized in $\gb$ read $c=\frac{\pi}{8G_5}\bigl(1-\tfrac{7}{2}\gb\bigr)$ and $b=\frac{\pi}{8G_5}\bigl(1-\tfrac{15}{2}\gb\bigr)$, which together with the expressions $W=0$ and $E=\frac{24\ddot{a}(k+\dot{a}^2)}{a^3}$ for the FLRW metric reduce the general expression \eq{GBcentralchargesdef} to \eq{CFTanomalyGBAdS}.

The results above generalize the expressions \eq{CFTenergypressureSAdS} and \eq{CFTanomalySAdS} obtained in the previous section to linear order in the Gauss-Bonnet parameter. They are believed to share qualitative features with the corresponding result for the $\mathcal{N}=4$ plasma including leading $1/\lambda$ corrections (here $\lambda=g_{_\text{YM}}^2N_c$ is the 't Hooft coupling). Once again one would like to emphasize that it follows naturally from our FLRW-like foliation \eq{BHmetricFLRW} of generic AdS spacetimes as a simple application to the AdS-Gauss-Bonnet black hole.

\subsection{AdS-Reissner-Nordström black hole}

We now turn to the case of a charged black hole in order to introduce a chemical potential for the dual plasma. This is the case, for instance, for the quark-gluon plasma of QCD which has a nonvanishing barion chemical potential. It is also interesting to illustrate how our procedure works when matter fields are present. The simplest bulk action includes a $U(1)$ gauge field minimally coupled to the Einstein-Hilbert action with a negative cosmological constant, namely
\begin{equation}
 S = \frac{1}{16\pi G_5}\int\ d^5x\sqrt{-g}\left[R+\frac{12}{L^2}-\frac{1}{4}F_{ab}F^{ab}\right]\ .
\end{equation}
An exact solution to the resulting Einstein and Maxwell equations is the AdS-Reissner-Nordström (AdSRN) black hole, a charged black hole whose metric can be cast in the standard black hole form \eq{BHmetricusual} with $f(r)$ and $\Sigma(r)$ given by (for simplicity we consider only the planar horizon case $k=0$)
\begin{equation}
 f(r) = \frac{r^2}{L^2}\left(1-\frac{M}{r^4}+\frac{Q^2}{r^6}\right)\qquad \Sigma(r)=\frac{r}{L}\ ,
\end{equation}
in addition to the nontrivial gauge field configuration
\begin{equation}\label{eq:RNAdSgaugefield}
 A_adx^a = \mu\left(1-\frac{r_h^2}{r^2}\right)dt\ .
\end{equation}
In the above we have conveniently expressed the solution in terms of four parameters, the mass $M$, charge $Q$, chemical potential $\mu$ and horizon radius $r_h$ (satisfying $f(r_h)=0$), but only two of them are independent parameters. For instance, $M$ and $Q$ can be eliminated in favor of $\mu$ and $r_h$ as
$$M = r_h^4+\frac{Q^2}{r_h^2}\ ,\qquad Q^2 = \frac{L^2\mu^2r_h^4}{3}\ .$$
The corresponding Hawking temperature is
\begin{equation}\label{eq:HawkingTRNAdS5k0}
 T_H = \frac{r_h}{\pi L^2}\left(1-\frac{L^2\mu^2}{6r_h^2}\right)\ .
\end{equation}

The AdSRN solution is believed to be holographically dual to a CFT plasma at temperature $T_H$ and chemical potential $\mu$ living at the AdS boundary $r=\infty$. Notice that there is a critical value for the chemical potential $\mu_c=\sqrt{6}r_h/L$ (correspondingly $Q_c=\sqrt{2}r_h^3$) where the temperature vanishes and the solution becomes extremal. 

The explicit form of the foliation \eq{BHmetricFLRW} for the AdSRN black hole, including the corresponding ($V,R$)-dependent gauge field configuration needed to support the metric\footnote{Namely, a gauge field configuration originally of the form $A_adx^a=\phi(r)dt$ (such as \eq{RNAdSgaugefield}) after the sequence of coordinate transformations $dt=dv-f(r)^{-1}dr$, $dv=a(V)^{-1}dV$ and finally $r=Ra(V)$ becomes $$A_adx^a = \phi(Ra)\left[\left(\frac{1}{a}-\frac{R\dot{a}}{f(Ra)}\right)dV-\frac{a}{f(Ra)}dR\right].$$}, takes the following form
\begin{eqnarray}\label{eq:RNAdSmetricFLRW}
 ds^2 &=& 2dVdR-\left[\frac{R^2}{L^2}\Big(1-\frac{M}{R^4a^4}+\frac{Q^2}{R^6a^6}\Big)-2R\frac{\dot{a}}{a}\right]dV^2+\frac{R^2a^2}{L^2}d\x^2\notag\\
 A_adx^a &=& \mu\left(1-\frac{r_h^2}{R^2a^2}\right)\left(\frac{1}{a}-\frac{L^2\dot{a}/Ra^2}{1-\frac{M}{R^4a^4}+\frac{Q^2}{R^6a^6}}\right)dV-\mu\left(1-\frac{r_h^2}{R^2a^2}\right)\frac{L^2/R^2a}{1-\frac{M}{R^4a^4}+\frac{Q^2}{R^6a^6}}dR,\qquad
\end{eqnarray}
where we have used $d\x^2$ to denote the spatial part $d\Omega_{k=0}^2$. As before, the local temperature $T(V)$ of the holographic dual nonequilibrium plasma is given by \eq{LocalT}.

It is interesting to notice that at the cosmological boundary $R=\infty$ what remains of the gauge field above is 
\begin{equation}
 A_\nu dx^\nu\big|_{_\text{bdry}} = \frac{\mu}{a}dV\ ,
\end{equation}
showing that the CFT plasma living in there has a time-dependent chemical potential $\tilde{\mu}(V)\equiv\frac{\mu}{a(V)}$.  Interestingly, by reversing the logic, one learns the important lesson that a CFT plasma subject to a time-dependent chemical potential $\tilde{\mu}(V)$ (what is sometimes referred to as a \emph{quench} in the chemical potential) experiences a non-equilibrium dynamics equivalent to a cosmological evolution with the scale factor being inversely related to the \lq\lq quench profile\rq\rq, i.e., $a(V)\sim1/\tilde{\mu}(V)$.

The AdSRN metric \eq{RNAdSmetricFLRW} parametrized by ($M,Q$) differs from the AdS-Schwarzschild solution \eq{SAdSmetricFLRW} only due to the presence of the charge $Q$, which comes with a subleading $1/R$ dependence and, thus, is expected to play a significant role only sufficiently deep inside the bulk spacetime. This suggests that the first terms in the Fefferman-Graham expansion of AdSRN near the cosmological boundary at $R=\infty$ should not differ from those obtained before. In fact, one can check explicitly that the transformation from ($V,R$) to FG coordinates ($\tau,z$) takes exactly the same form as in \eq{VRtoFGsolSAdS} (with $Q$ only beginning to affect at $\mathcal{O}(z^7)$ in $V(\tau,z)$ and $\mathcal{O}(z^5)$ in $R(\tau,z)$) and, consequently, that the first few FG metric coefficients are the same as in \eq{FGcoeffsSAdS} (with $Q$ only beginning its influence at $\mathcal{O}(z^6)$ in both $g_{\tau\tau}$ and $g_{ij}$).\footnote{Of course we may want to parametrize the AdSRN metric using not ($M,Q$) but instead ($r_h,Q$), for instance. In this case the FG coefficients are still going to be given by the AdS-Schwarzschild expressions \eq{FGcoeffsSAdS} with $M$ now expressed in terms of ($r_h,Q$) as $M=r_h^4+\frac{Q^2}{r_h^2}$ (and obviously the statement about $Q$ only starting to affect the expansion at higher orders must be forgotten).}
The gauge field appearing in \eq{RNAdSmetricFLRW}, on the other hand, has a nontrivial FG expansion that is readily found to be
\begin{eqnarray}
 A_adx^a &=& \mu\bigg[\frac{1}{a}-\frac{a\ddot{a}-2\dot{a}^2+2r_h^2}{2a^3}z^2
 +\cdots\bigg]d\tau -\mu\bigg[\frac{\dot{a}}{a^2}z+\frac{\dot{a}\left(\dot{a}^2-2r_h^2\right)}{2a^4}z^3
+\cdots\bigg]dz\ .
\end{eqnarray}

The holographic renormalization for the Einstein-Maxwell system can be found, e.g., in \cite{Sahoo:2010sp}. The expressions for the renormalized stress tensor and $U(1)$ conserved current of the dual CFT are generically given in terms of the FG coefficients as\footnote{Up to scheme dependent terms that do not contribute to the conformal anomaly and can be removed by additional counterterms (see \cite{Sahoo:2010sp} for details).}
\begin{eqnarray}\label{eq:TmunuCFTeinsteinmaxwell}
\bigl\langle T_{\mu\nu}\bigr\rangle &=& \frac{1}{ 4\pi G_5}\left\{g^{(4)}_{\mu\nu}-\frac{1}{2}g^{(2)\sigma}_{\ \mu}g^{(2)}_{\sigma\nu}+\frac{1}{4}\bigl(g^{(2)\sigma}_{\ \sigma}\bigr) g^{(2)}_{\mu\nu}-\frac{1}{8}\bigl[\bigl(g^{(2)\sigma}_{\ \sigma}\bigr)^2g^{(2)}_{\sigma\rho}g^{(2)\rho\sigma}\bigr]g^{(0)}_{\mu\nu}+\frac{1}{48}\bigl(F^{(0)}_{\sigma\rho}F^{(0)\sigma\rho}\bigr)g^{(0)}_{\mu\nu}\right\}\notag\\
\bigl\langle J^{\mu}\bigr\rangle &=& \frac{1}{8\pi G_5}\bigl(A^{(2)}_\nu+B^{(2)}_\nu\bigr)g^{(0)\nu\mu} \ ,
\end{eqnarray}
where $A^{(2)}_\nu$ and $B^{(2)}_\nu$ are the second order coefficients appearing in the FG expansion of the bulk gauge field, i.e., 
$$A_{\nu}(z,x)=A^{(0)}_\nu(x)+z^2\bigl[A^{(2)}_\nu(x)+B^{(2)}_\nu(x)\log z^2\bigr]+\cdots\ .$$
Note that the expression for $\langle T_{\mu\nu}\rangle$ is almost the same as the one for pure Einstein gravity in the bulk, \eq{TmunuCFTeinstein}, the exception being the extra contribution due to the gauge field in the last term. Inserting our FG expansion constructed above gives the following result for the energy density, pressure, and charge density $\mathcal{Q}\equiv\bigl\langle J^{\tau}\bigr\rangle$ of a $U(1)$-charged plasma in a FLRW spacetime
\begin{eqnarray}\label{eq:CFTenergypressureRNAdS}
\mathcal{E} &=& \frac{3\dot{a}^4+12r_h^2(r_h^2+\frac{1}{3}\mu^2)}{64\pi G_5 a^4} \nonumber\\
\mathcal{P} &=& \frac{\dot{a}^4+4r_h^2(r_h^2+\frac{1}{3}\mu^2)-4a\ddot{a}\dot{a}^2}{64\pi G_5 a^4}\nonumber\\
\mathcal{Q} &=& \frac{\mu\bigl(2r_h^2+2\dot{a}^2+a\ddot{a}\bigr)}{16\pi G_5a^3}\ .
\end{eqnarray}
Once again this can be put entirely in 4d CFT language using $G_5=\frac{\pi}{2N_c^2}$, eliminating $r_h$ in favor of the local temperature $T(V)$ as $r_h=\frac{1}{2}\pi aT\bigl(1+\sqrt{1+2\mu^2/3\pi^2T^2}\bigr)$ and, finally, writing $\mu=\tilde{\mu}a$, since the chemical potential associated to the expanding plasma is the time-dependent one $\tilde{\mu}=\frac{\mu}{a}$ instead of $\mu$, as discussed above. The resulting expressions are lengthy and no more instructive than \eq{CFTenergypressureRNAdS}, so we do not show them explicitly here. We just point out, as a sanity check, that for $\mu=0$ the result \eq{CFTenergypressureSAdSin4dlanguagek=0} is sucessfully recovered.

The conformal anomaly in this case is the same as in \eq{CFTanomalySAdS}, i.e., there is no contribution from the chemical potential to the anomaly. The $U(1)$ gauge field usually contributes a term proportional to $\bigl(F^{(0)}_{\mu\nu}\bigr)^2$ (see \cite{Sahoo:2010sp}), but for our solution this is zero since at the boundary we only have $A^{(0)}_\nu dx^\nu=\frac{\mu}{a}d\tau$ (hence $F^{(0)}_{\mu\nu}=0$). It is straightforward to check from \eq{CFTenergypressureRNAdS} that 
\begin{eqnarray}
 \nabla_\mu\langle T^{\mu\nu}\rangle &=& 0\notag\\
 \nabla_\mu\langle J^{\mu}\rangle &=& \frac{\mu}{2a^3}\left(-3\dot{a}\ddot{a}+a\dddot{a}\right)\ ,
\end{eqnarray}
i.e., the CFT energy-momentum tensor is conserved while the $U(1)$ current is not. This is a direct consequence of the dynamical chemical potential experienced by the plasma and should not come as a surprise.

\section{Conclusions}\label{sec:AdSFLRWconclusions}

We have introduced a new slicing of AdS black holes such that a non-standard notion of conformal boundary with a FLRW metric can be defined. 
The construction is based on the use of Eddington-Finkelstein coordinates instead of early approaches involving Fefferman-Graham coordinates, a fact that makes the task tremendously simpler and applicable to a large class of AdS black holes including eventual supporting matter fields. It also provides a good perspective into the numerical study of expanding plasmas in holography using the characteristic formulation of the Einstein equations in AdS \cite{Chesler:2013lia}, for which the use of EF coordinates is determinant. A Fefferman-Graham expansion near the new \lq\lq cosmological\rq\rq~boundary can be easily constructed to any desired order for the FLRW-foliated metric, which leads to the renormalized stress tensor of the dual expanding CFT plasma upon the assumption that the standard holographic renormalization procedure is still applicable. In particular, the results of \cite{Apostolopoulos:2008ru} are consistently recovered as a simple application to the AdS-Schwarzschild solution and then generalized to include a second central charge (using the AdS-Gauss-Bonnet black hole) or a nonvanishing chemical potential (using the AdS-Reissner-Nordström solution) for the dual CFT plasma.

The new dynamical foliation elucidates the procedure carried out in \cite{Buchel:2016cbj} by clarifying the assumptions involved and the background solution on which the perturbative study of the expanding $\mathcal{N}=2^*$ plasma close to conformality relies. It also provides, as a by-product of the application to the AdS-Reissner-Nordström solution, the interesting lesson that the nonequilibrium dynamics of a CFT plasma subject to a quench $\tilde{\mu}(t)$ in the chemical potential resembles a cosmological evolution with the scale factor $a(t)$ being inversely related to the quench profile, $a(t)\sim \tilde{\mu}(t)^{-1}$. A similar conclusion can be drawn for a wider class of quenches of CFTs by applying our slicing to the corresponding dual static hairy black hole solution, since the (time-independent) non-normalizable mode associated with static matter field configurations naturally acquires time dependence in the new ($V,R$) coordinates.

The proposal also gives a novel tool to analytically explore properties of expanding plasmas that have not yet been explored. This involves, for instance, a study of the time evolution of nonlocal observables with a known holographic dual gravity description, such as higher-point correlators, Wilson loops and the entanglement entropy of spatial subsystems. However, we choose to postpone this to a subsequent work.

\subsection*{Acknowledgements}
I am very grateful to Elcio Abdalla, Jorge Noronha, Ricardo Landim, Viktor Jahnke and Leonardo Werneck for useful comments and discussions. I also thank Conselho Nacional de Desenvolvimento Científico e Tecnológico (CNPq) for financial support.

\bibliographystyle{JHEP}
\bibliography{myrefAdSFLRW,myrefBooks,myref}

\end{document}